\documentclass[twocolumn,showpacs,preprintnumbers,amsmath,amssymb]{revtex4}
\usepackage{graphicx}% Include figure files
\usepackage{dcolumn}% Align table columns on decimal point
\usepackage{bm}% bold math

\begin{document}

%\preprint{APS/123-QED}

\title{Effective transition operators for resonant inelastic X-ray scattering}

\author{Michel van Veenendaal}
\affiliation{%
Dept. of Physics, Northern Illinois University,
De Kalb, Illinois 60115\\ 
Advanced Photon Source,
  Argonne National Laboratory, 9700 South Cass Avenue, Argonne,
Illinois 60439}%

%\affiliation{

\date{\today}% It is always \today, today,

\begin{abstract}
Effective symmetry-based transition operators for resonant inelastic X-ray scattering (RIXS) are derived that show how the scattering between different states depends on the polarization of the incoming and outgoing X-rays. In spherical symmetry, the effective operators can be rewritten in terms of spin operators, although the expressions depend on the nature of the ground state. For lower symmetries, the combined action of the crystal field and the spin-orbit interaction breaks up the spin space and spin operators are no longer appropiate operators. By taking  iridium compounds as an example, it is demonstrated that  effective scattering operators can still be obtained. These effective transition operators facilitate our understanding how RIXS couples to elementary excitations.
\end{abstract}

\maketitle

Collective excitations play an important role in the electronic and magnetic properties of materials. A typical example is a magnon, a spin-flip excitation that disperses through the solid via the exchange interactions. These excitations are typically studied with inelastic neutron scattering where magnons are created through the interaction of the spin of the neutron with the magnetic moments of the material. However, magnon excitations can also be studied via the inelastic scattering of X-rays \cite{DeGroot,Braicovich,Schlappa,BrinkEPL,Hill}. The creation of electronic excitations can be enhanced by tuning the X-ray energy to a particular resonance where an electron from a deep-lying core level is excited into the valence states. Using this technique, known as resonant inelastic X-ray scattering (RIXS) \cite{AmentRMP}, magnon dispersions throughout the Brillouin zone have been mapped for copper-oxide based materials \cite{Braicovich,Schlappa}. More recent experiments on iridates \cite{Ishii,Kim} provide an excellent example of how RIXS and inelastic neutron scattering can complement each other. Iridates have come to the forefront recently \cite{KimScience,Jackeli,LagunaMarco,Ament} as a prototypical example of a system where the orbital and spin of the valence electrons are entangled by the strong spin-orbit interaction. However, the small size of iridium-oxide single-crystals  and its large atomic weight significantly hamper the study of these materials using neutrons. Using RIXS, Kim {\it et al.} \cite{Kim} were able to map a magnon dispersion with a width of about 0.2 eV in Sr$_2$IrO$_4$. In addition, excitations were observed \cite{Ishii,Kim} with an energy of 0.5-0.8 eV which were assigned to entangled spin-orbital waves. 

One of the attractive features of inelastic neutron scattering is that its cross section is due to a comparatively simple dipole interaction between the neutrons and the magnetic moments in the solid. The RIXS spectral intensity, on the other hand, is due to a complex process, where after absorption of the incoming X-ray, the system is in a complicated intermediate state dominated by strong interactions with the core hole. However, it is exactly the large core-hole spin-orbit coupling and the many-body effects between the core hole and the valence shell that make spin flips possible. Subsequently, the final states are reached via the radiative decay of the core hole. Analytical evaluation of the resonant scattering amplitude \cite{Luo,MvVRIXS,JvdB,MvVdd} can provide more insight into the spectal weights of the final states, but generally require approximations of the intermediate-state propagator. However, when studying the momentum and polarization dependence of magnons and orbitons, one is less interested in the intensity compared to features of different symmetry, but more in the relative strength of the components of a particular symmetry. For example, for magnon excitations, it is important to understand how  the polarization conditions affect the ratio between spin conserving and spin-flip transitions.
In this Letter, it is shown that these transitions can be expressed in relatively simple operators. Although the operators can often  be expressed in terms of spin operators \cite{Haverkort}, this result is not general and depends on the nature of the ground state. The relative strength of the operators can be tuned by the polarization vectors of the incoming and outgoing X-rays.

The RIXS intensity is given by the Kramers-Heisenberg equation \cite{AmentRMP},
\begin{eqnarray}
I(\omega,\omega',{\bm \epsilon},{\bm \epsilon}')=
\sum_f|\langle f| {\cal F}(\omega,{\bm \epsilon},{\bm \epsilon}') |i\rangle|^2 \delta(E_f+\hbar \omega'-E_i +\hbar \omega ) \nonumber
\end{eqnarray}
 where the energy loss $\hbar( \omega- \omega')$ of the photon equals the difference $E_f-E_i$ between final and initial energies  of the system; ${\bm \epsilon}$ and ${\bm \epsilon}'$ are the polarization vectors of the incoming and outgoing X-ray photons, respectively. The scattering amplitude consists of dipole operators $ {\bf D}$ describing the absorption and radiative decay and an intermediate-state propagator:
\begin{eqnarray}
{\cal F}(\omega,{\bm \epsilon},{\bm \epsilon}')=  {\bm \epsilon}'^*  \cdot {\bf D} \frac{1}{\hbar \omega-H +i\Gamma} {\bm \epsilon} \cdot {\bf D} 
\end{eqnarray}
where $H$ is the Hamiltonian of the system and $\Gamma$ the intermediate-state lifetime broadening. In order to treat systems with different local symmetries, we define $|m\rangle=|\Gamma_m \gamma_m\rangle$ for $m=i,f$, where $\Gamma_m$ (not to be confused with the lifetime broadening) is the total symmetry and $\gamma_m$ its component. For example, for a divalent nickel ion in octahedral symmetry ($O_h$), the total symmetry is $\Gamma=^3A_2$ and the components $\gamma=1,0,-1$ are the three spin states of the triplet. After recoupling \cite{Hannon,AmentRMP} , the scattering amplitude, 
\begin{eqnarray}
{\cal F}(\omega,{\bm \epsilon},{\bm \epsilon}')=\sum_{k=0}^2 \sum_q a_k T^{k*}_q ({\bm \epsilon},{\bm \epsilon}') F^k_q (\omega) ,
\end{eqnarray}
can be separated  into a simple numerical factor $a_k=(2k+1)n^2_{11k}$ \cite{normalization}, a term $T^{k}_q({\bm \epsilon},{\bm \epsilon}')=[{\bm \epsilon}'^*,{\bm \epsilon}]^k_q$  describing the dependence on the polarization vectors as a tensor product \cite{normalization}, where $q=\Gamma\gamma$ is the symmetry of the transition operator,  and an effective scattering operator $F^k_q$.
Using the Wigner-Eckart theorem \cite{Butler}, we split the matrix element of $F^k_q$ into two terms
\begin{eqnarray}
\langle f| F^k_q (\omega)|i\rangle= (\Gamma_f ||F^k(\omega) || \Gamma_i)n_{ifk} \langle f|W_q |i\rangle,
\end{eqnarray}
where the reduced matrix element $(\Gamma_f ||F^k(\omega) || \Gamma_i)$ describes the strength of the transition between the manifolds of symmetry $\Gamma_i$ and $\Gamma_f$. The second term can be written as an effective operator
\begin{eqnarray}
W_q =\sum_{if} |f\rangle \langle f|qi\rangle \langle i| n_{ifk}^{-1},
\label{Wq}
\end{eqnarray}
where $n_{ifk}$ are factors that are introduced to obtain more nicely defined operators. The term $\langle f|qi\rangle$ is a coefficient that gives the coupling between the initial state $i$ and the final state $f$ via a tensor with symmetry $q$. It is important to realize that all the dependence on the intermediate states is contained in the reduced matrix elements that only depend on the total symmetries $\Gamma_i$ and $\Gamma_f$. The polarization dependence of the  intensities of the different components of a particular manifold only depends on the effective operators $W_q$. Although the above result is exact, it is not directly clear that the operators $W_q$ are of any practical use since they depend on coupling constants $\langle f|qi\rangle$ which are matrix elements between complex initial and final states. In the following, it is demonstrated that the $W_q$ can be expressed as relatively simple operators. 

Since the nature of the operators $W_q$ depends on the symmetry of the problem, we focus on some specific situations. Let us first consider spherical symmetry. Typical systems are rare-earth ions, where generally the orbital moment ${\bf L}$ and spin ${\bf S}$ are coupled to a total moment ${\bf J}$ via the spin-orbit coupling. The total symmetry of the state is characterized by the total momentum ($\Gamma_m\rightarrow J$); the components are given by the projection ($\gamma_m\rightarrow M_J$). This gives $|i\rangle=|JM_J\rangle$ and $|f\rangle=|J'M'_J\rangle$. In a magnetic field along the positive $z$ axis, the ground state has $M_J=-J$. For the scattering operators, all the components of the scattering term of rank $k$ belong to the same representation and $\Gamma\gamma\rightarrow kq$ with $q=k,k-1,\cdots,-k$. The coupling constants in Eq. (\ref{Wq}) are directly related \cite{CondonShortley} to the Clebsch-Gordan coefficients $\langle J'M_J'|kq JM_J\rangle$. Obviously, well-defined effective operators are the components $J^{(1)}_q=J_q$ of the total angular momentum ${\bf J}$, whose matrix elements are proportional to $\langle J'M_J'|1q JM_J\rangle$, and combinations thereof, such as $J_qJ_{q'}$, that give rise to higher-order terms $J^{(k)}_q$. Expressing the operators $W_{kq}$ in terms of spin requires some more care, since the ground state is not directly an eigenstate of the spin. Let us consider excitations within the ground-state manifold characterized by $J$. If the ion is coupled via exchange interactions to other magnetic ions in the lattice, these excitations would lead to spin waves. For these states, we can define general spin operators of rank $k$ via
\begin{eqnarray}
\sigma^{(k)}_q =\sum_{S_z,S'_z} |SS'_z\rangle \langle SS'_z|kq SS_z\rangle
n^{-1}_{SSk} \langle  SS_z| ,
\label{spin}
\end{eqnarray}
where $S$ is the spin. ${\bm \sigma}^{(0)}$ is the identity operator; the Pauli spin matrix is given by ${\bm \sigma}^{(1)}\mathopen{=}{\bf S}/S$; the spin operator of rank 2 is normalized by ${\bm \sigma}^{(2)}\mathopen{=}{\bf S}^{(2)}/S(2S\mathopen{-}1)$. Note that the maximum change in spin is given by $k$.  In spherical symmetry, we can relate the spin operator to the total angular momentum operators $J_q$ via
\begin{eqnarray}
\langle JM'_J| \sigma^{(k)}_q |JM_J\rangle =
A_{LSJk} \langle JM'_J| J^{(k)}_q |JM_J\rangle ,
\end{eqnarray}
where the $A_{LSJk}$ gives the details of the coupling between the orbital moment $L$ and the spin $S$ in the ground state 
\begin{eqnarray}
A_{LSJk} =\sqrt{(2J+1)(2S+1)} \frac{n_{JJk}}{n_{SSk}} 
\left \{
\begin{array}{ccc}
J & J& k\\
S & S & L
\end{array}
\right \},
\end{eqnarray}
where the bracketed term is a $6j$ symbol.  When $J=L+S$, one has $A_{LSJz}=1$ and the "spin-only" result of Haverkort \cite{Haverkort} is reproduced. However, when the spin is antiparallel to the orbital moment, the weighting of the different ${\bm \sigma}^{(k)}$ to the resonant X-ray scattering amplitude changes. For example, for $J=L-S$ with $L>S$,  $A_{LSJ1}=-J/(J+1)$ and $A_{LSJ2}=J(2J-1)/(J+1)(2J+3)$. Therefore, in spherical symmetry, the inelastic scattering amplitude can be expressed in terms of spin operators, since the spin operator is a proper spherical tensor However the expressions depend on the coupling between the orbital and spin moment in the ground state.

An interesting feature of scattering  within the ground-state manifold is that the reduced matrix element can be expressed in terms of the elastic scattering amplitude
\begin{eqnarray}
{\cal F}^k_{\rm el.}(\omega) =\langle J,-J| F^k_{0}(\omega)|J,-J\rangle =(J || F^k (\omega)|| J) n_{JJk}. \nonumber
\end{eqnarray}
which is directly related to the X-ray absorption, which is  proportional to $-{\rm Im}[{\cal F}^k_{\rm el.}(\omega)]$. For the ground-state manifold with total angular momentum $J$, the inelastic scattering from $M_J$ to $M'_J$, of importance for magnetic excitations can be written as
\begin{eqnarray}
{\cal F}(\omega,{\bm \epsilon},{\bm \epsilon}')=\sum_{k=0}^2  \frac{a_k}{A_{LSJk}}{\cal F}^k_{\rm el.}(\omega) [{\bm \epsilon}'^*,{\bm \epsilon}]^k \cdot {\bm \sigma}^{(k)}_M
\end{eqnarray}
where the polarization dependence is directly coupled to the effective operator.

\begin{figure}[t]
 \includegraphics[width=0.5\columnwidth]{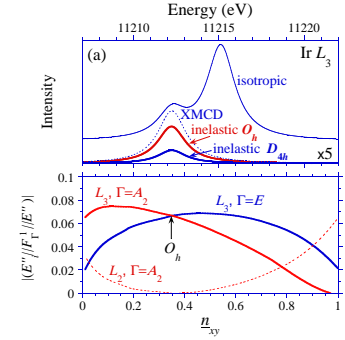}
\caption{ (a) The imaginary part of the  X-ray scattering amplitudes [$-(1/\pi) {\rm Im}\langle E''s'_z |F^k_{\Gamma\gamma} |E''\frac{1}{2}\rangle $] for an Ir$^{4+}$ ion ($t_{2g}^5$). Shown are the isotropic X-ray absorption ($k=0$,  $s'_z=\frac{1}{2}$), the x-ray magnetic circular dichroism (XMCD) ($k=1$,  $s'_z=\frac{1}{2}$), and the inelastic spin-flip scattering amplitude ($k=1$,  $s'_z=-\frac{1}{2}$) in planar symmetry ($D_{4h}$) in the limit that the hole density in the $xy$ orbital ${\underline n}_{xy}$ goes to zero. For comparison, the inelastic spin-flip scattering in octahedral symmetry is shown.  (b) The reduced matrix element for the $k=1$ transitions as a function of the hole density ${\underline n}_{xy}$ in the $xy$ orbital. The terms with and without spin flip are given by $\Gamma=E,A_2$, respectively. There are no spin flip terms at the $L_2$ edge.
}
\label{iridium} 
\end{figure}

The correspondence between the effective operators and the local spin operators holds in many cases, but fails for lower symmetries where the combined action of the crystal-field and the spin-orbit coupling can effectively break up the spin space. A representative example is a low-spin tetravalent iridium ion. Iridates have received significant attention recently \cite{KimScience,Kim,Jackeli,Ishii,LagunaMarco,Ament} since in these systems, despite a large bandwidth, the spin-orbit interaction is not quenched. This can lead to new kinds of topological orderings. Ir$^{4+}$ has a $t^5_{2g}$ configuration ($^2T_2$ in $O_h$).  Orbital and spin space are coupled via the spin-orbit interaction and the  $^2T_2$ symmetry splits into $E''$ and $U'$ symmetries, which are two- and four-fold degenerate, respectively. For an iridium ion, the $E''$ representation is the lowest in energy. This still holds in planar symmetry ($D_{4h}$) and the wavefunction for the $t_{2g}$ hole in the  ground state can be written as \cite{Jackeli}
\begin{eqnarray}
|E'',\pm \frac{1}{2}\rangle=\sin\alpha |0,\pm \frac{1}{2}\rangle-\cos \alpha |\pm 1, \mp\frac{1}{2}\rangle,
\label{IrGS}
\end{eqnarray}
where $|0\rangle=-i|xy\rangle$ and $|\pm 1\rangle=\frac{1}{\sqrt{2}} (-|zx\rangle\pm i|yz\rangle$). Figure \ref{iridium}(a) shows the x-ray absorption and x-ray magnetic circular dichroism at the $L_3$ edge. The spectra are calculated using standard multiplet codes \cite{LagunaMarco}, including a cubic crystal field of 3 eV. The symmetry is further lowered by applying a planar crystal field, which also determines the relative occupation of the $xy$ and $yz/zx$ orbitals. The spectra are relatively simple: the main line is due to the four empty $e_g$ states; at the low-energy side, there is intensity due to the empty $t_{2g}$ state, which shows a dichroic signal.  Let us consider the scattering for $k=1$ between the different $S_z$ components  of the ground-state manifold $|E''S_z\rangle$ with $S_z=\pm \frac{1}{2}$. In planar symmetry, the three components of the $k=1$ transition operator  are $\Gamma\gamma=A_2; E,1; E,-1$ which are given by $W_{A_2}=-\frac{1}{3}{\overline \sigma}^{(1)}_z=-\frac{2}{3}{\overline S}_z$ and $W_{E,\pm 1}=\frac{1}{3}{\overline \sigma}^{(1)}_{i}=\frac{2}{3}{\overline S}_{x/y}$  where the spin operators are given in Eq. (\ref{spin}) using the equivalence between $E''$ and $S=\frac{1}{2}$. Although the operators are relatively simple, it is important to note that they are effective spin operators (indicated by the overline in ${\overline S}_{i}$) between the two components of the $E''$ ground state and not real spin operators. This becomes obvious when considering  the limit $\alpha\rightarrow 0$ in Eq. (\ref{IrGS}) where the low-energy manifold is given by $|E'',\pm \frac{1}{2}\rangle= |\pm 1, \mp\frac{1}{2}\rangle$. The matrix elements of the spin operator are given by $\langle E'' S'_z|S_i |E''S_z \rangle= -S_z \delta_{S_z,S'_z}\delta_{i,z}$. Therefore, from the point of view of the (real) spin, the $|E'',\pm\frac{1}{2}\rangle$ manifold is an effective Ising spin due to the fact that a spin flip must be accompanied by a change in orbital. This implies that if the inelastic x-ray scattering amplitude would be described solely in terms of spin operators, magnon excitations between $|E''\frac{1}{2}\rangle$ and $|E'',-\frac{1}{2}\rangle$ would not be allowed which clearly contradicts the finite spin-flip scattering found numerically, see Fig. \ref{iridium}(a).

For excitations within the manifold with the lowest energy, the $k=1$ transition operator is proportional to
\begin{eqnarray}
\sum_{i=x,y}( E''|| F^1_{E}(\omega) || E'')P_i {\overline \sigma}^{(1)}_i
-( E''|| F^1_{A_2}(\omega) || E'')P_z {\overline \sigma}^{(1)}_z \nonumber
\end{eqnarray}
with $P_i\equiv T^1_i({\bm \varepsilon},{\bm \varepsilon}')= ({\bm \varepsilon}'^*\times {\bm \varepsilon})_i$ with $i=x,y,z$.  Within planar symmetry, the $S_z$ and spin-flip terms have a different symmetry and their relative weight therefore depends on the frequency-dependent reduced matrix element. Since the spin excitations occur in the ground-state manifold, we can express the inelastic scattering amplitudes in terms of the elastic one. For the $A_2$ representation, $( E''|| F^1_{A_2}(\omega) || E'')=F^1_{A_2,{\rm el.}}(\omega)/n_{\frac{1}{2}\frac{1}{2}1}$, where  $-{\rm Im}[F^1_{A_2,{\rm el.}}(\omega)]$ is the XMCD with the magnetic or exchange fields in the $z$ direction. However, the coupling coefficient is zero for $E$ symmetry in this configuration. It is nonzero when the magnetization is in the plane, $( E''|| F^1_{A_2}(\omega) || E'')=F^1_{Ex,{\rm el.}}(\omega)/n_{\frac{1}{2}\frac{1}{2}1}$ where ${\rm Im}[F^1_{Ex,{\rm el.}}(\omega)]$ is the XMCD measured along the $x$ direction. The values of the reduced matrix elements are shown in Fig. \ref{iridium}(b) as a function of the number of holes ${\underline n}_{xy}$ in the $xy$ orbital.  Although there are no spin-flip terms at the $L_2$ edge, the spin-conserving and spin-flip terms are of comparable order of magnitude at the $L_3$ edge. When there is no planar distortion, (${\underline n}_{xy}\cong 0.35$), all directions become equivalent and the scattering amplitude becomes proportional to ${\bf P}\cdot {\overline {\bm \sigma}}^{(1)}$. Since there are no spin flips at the $L_2$ edge, this also implies that the reduced matrix element for ${\overline \sigma}^{(1)}_z$ must be zero, see Fig. \ref{iridium}(b).

The approach described here is not restricted to excitations within the ground-state manifold, but can also describe transitions between different manifolds. Let us consider again the situation for an Ir$^{4+}$ ion and take octahedral symmetry [$\cos^2\theta=\frac{2}{3}$ in Eq. (\ref{IrGS})]. The wavefunctions for $U'$ symmetry are $|U',\pm\frac{3}{2}\rangle=\mp |\pm 1,\pm \frac{1}{2}\rangle$ and $|U',\pm\frac{1}{2}\rangle=\pm  \sqrt{\frac{2}{3}}|0,\pm \frac{1}{2}\rangle\pm \sqrt{\frac{1}{3}}|\pm 1,\mp \frac{1}{2}\rangle$. Let us consider the case of $k=1$, which in octahedral symmetry corresponds to the $T_1$ representation with components $q=1,0,-1$. The effective transition operator given by
\begin{eqnarray}
W_{T_1q}\sim \sum_{S_z} |U', S_z\mathopen{+}q\rangle\langle U',S_z\mathopen{+}q|T_1 q,E''S_z\rangle 
 \langle E''S_z|
\end{eqnarray}
brings the system from the ground state with $S_z=\pm \frac{1}{2}$ into the four-fold degenerate final states. The matrix elements of $W_{T_1q}$ can be directly related to a Clebsch-Gordan coefficient
\begin{eqnarray}
\langle U',S_z\mathopen{+}q|T_1 q,E''S_z\rangle
\sim\frac{\langle \frac{3}{2},S_z\mathopen{+}q|1q,\frac{1}{2}S_z\rangle}
{2n_{\frac{1}{2}1\frac{3}{2}}}=\sqrt{\frac{|S_z+q|}{1+|q|}}.
\nonumber
\end{eqnarray}
and the operator can be written as an effective total angular momentum ${\overline {\bf J}}$. The contributions of $W_{T_1q}$ to the scattering amplitude can be varied via the polarization dependence giving a weight $-(P_x+iP_y)/\sqrt{2},(P_x+iP_y)/\sqrt{2},P_z$ for $q=1,0,-1$, respectively. This allows one to probe different components of the orbiton state.

In conclusion, it has been shown that when considering the polarization dependence of excitation of a particular symmetry, such as magnon or phonons, relatively simple effective transition operators can be derived that depend only on the local symmetry. The coupling of these operators to the polarization dependence allow the tuning of excitations into the different components of the final state (for example, transitions into states with and without spin flip). 

We acknowledge useful discussions with Maurits Haverkort and Jeroen van den Brink. This work was supported by NIU's Institute for Nanoscience, Engineering, and Technology. This work was  supported by grant  DE-FG02-03ER46097 and the RIXS collaboration of the Computational Materials Science Network under grant number DE-FG02-
08ER46540 by the Division of Materials Science and Engineering, Office of Basic Energy Sciences, US Department of Energy (DOE). Work at Argonne National Laboratory was supported by the U.S. DOE, Office of Science, Office of Basic Energy Sciences (BES), under contract DE-AC02-06CH11357.


\begin{thebibliography}{99}
\bibitem{normalization}
The normalization constants are defined such that more nicely defined quantities can be defined. For spherical symmetry, we define
\begin{eqnarray}
n_{ll'L}=\frac{(l+l')!}{(l+l'-L)!L!}
\sqrt{\frac{(g-2l)!(g-2l')!(g-2L)!}{(g+1)!}},
\end{eqnarray}
with $g=l+l'+L$. These constants are used to define tensor products 
\begin{eqnarray}
[U^{l'},V^{l}]^L_M&=&\sum_{mm'} U^{l'}_{m'} (-1)^mV^{l}_{-m} \\
&& \times
(-1)^{l'-m'} \left ( 
\begin{array}{ccc}
l' & L & l \\
-m' & M & m 
\end{array}
\right )
 n^{-1}_{ll'L} ,
\end{eqnarray}
where the term in parentheses is a $3j$ symbol. This definition leads to vector products and operators that do not contain square roots resulting from the normalization to unity of the $3j$ symbol.
\bibitem{CondonShortley} The coupling constants are expressed in $W_{kM}$ are expressed in terms of $3j$ symbols and therefore differ by a factor $\sqrt{2J+1}(-1)^{J+J'-k}$ from the Clebsch-Gordan coefficients.
\end{thebibliography}
\end{document}